\documentclass[twocolumn, superscriptaddress, floatfix, showpacs, aps, prb,10pt]{revtex4-1}
\pdfoutput=1
\usepackage{graphicx}
\usepackage{amsmath}
\usepackage[colorlinks=true, citecolor=blue, urlcolor=blue]{hyperref}
\usepackage{amssymb}
\usepackage{bm}
\usepackage{color}
\usepackage{array}
\usepackage{booktabs}
\usepackage{multirow}
\usepackage{wasysym}
\usepackage[normalem]{ulem}

\newcommand{\pd}{{\phantom\dag}}

\begin{document}

\title{Topological crystalline insulators from stacked graphene layers}

\author{Sanjib Kumar Das}
\affiliation{Institute for Theoretical Solid State Physics, IFW Dresden, Helmholtzstrasse 20, 01069 Dresden, Germany}
\author{Binghai Yan}
\affiliation{Department of Condensed Matter Physics, Weizmann Institute of Science, Rehovot, 7610001, Israel}
\author{Jeroen van den Brink}
\affiliation
{Institute for Theoretical Solid State Physics, IFW Dresden, Helmholtzstrasse 20, 01069 Dresden, Germany}
\affiliation{Department of Physics, Technical University Dresden, 01062 Dresden, Germany}
\author{Ion Cosma Fulga}
\affiliation{Institute for Theoretical Solid State Physics, IFW Dresden, Helmholtzstrasse 20, 01069 Dresden, Germany}

\date{\today}
\begin{abstract}
In principle the stacking of different two-dimensional (2D) materials allows the construction of 3D systems with entirely new electronic properties.
Here we
propose to realize topological crystalline insulators (TCI) protected by mirror symmetry in heterostructures consisting of graphene monolayers separated by two-dimensional polar spacers. 
The polar spacers are arranged such that they can induce an alternating doping and/or spin-orbit coupling in the adjacent graphene sheets.
When spin-orbit coupling dominates, the non-trivial phase arises due to the fact that each graphene sheet enters a quantum spin-Hall phase. Instead, when the graphene layers are electron and hole doped in an alternating fashion, a uniform magnetic field leads to the formation of quantum Hall phases with opposite Chern numbers.
It thus has the remarkable property that unlike previously proposed and observed TCIs, 
the non-trivial topology is generated by an external time-reversal breaking perturbation.
\end{abstract}
\maketitle

\section{Introduction}
\label{sec:intro}

The foundation of topology in condensed matter physics was first laid by the experimental discovery of the integer quantum Hall effect\cite{Klitzing1980} and the subsequent theoretical work on quantized Hall conductances in two-dimensional (2D) periodic potentials.\cite{Thouless1982} However, the different ways in which topology can manifest in crystals were mostly unexplored until the prediction of the quantum spin Hall effect in graphene,\cite{Kane2005a, Kane2005b} which was termed a $\mathbb{Z}_2$ topological insulator (TI). Soon after this, the quantum spin Hall effect and the associated topological phase transition were experimentally observed in HgTe quantum wells.\cite{Bernevig2006, Konig2007} In the following years, the study of topological phases of matter has led to numerous rich discoveries in various condensed matter systems.\cite{Hasan2010, Qi2011, Fu2007}

Topological insulators are defined as having a gapped bulk, but hosting gapless, anomalous states on their boundaries, states which are protected by the symmetry of the system. Depending on the nature of the symmetry, topologically non-trivial phases are characterized by different integers, called topological invariants. A change in the value of these invariants marks a transition to a topologically different phase, one hosting either a different number of boundary states, or boundary states of a different chirality.
A systematic classification of which types of topological phases are possible was first carried out in the case of fundamental symmetries: time-reversal (TRS), particle-hole, and chiral symmetry.\cite{Schnyder2008, Kitaev2009}
Apart from these fundamental symmetries however, spatial symmetries can also give rise to topological insulating phases in materials. The latter are called weak topological insulators in the case of lattice translations,\cite{Fu2007} and topological crystalline insulators (TCI)\cite{Fu2011,Hsieh2012,Ando2015} for symmetries such as mirror, rotation, or glide.
Recently, the experimental discovery of mirror symmetry protected TCIs in the SnTe material class has made a tremendous impact in this field of research.\cite{Tanaka2012, Dziawa2012, Xu2012} There have been many works reported in the literature, classifying TCI based on their lattice symmetries,\cite{Slager2012, Chiu2013, Morimoto2013, Jadaun2013, Benalcazar2014, Fang2015, Diez2015}
proposing new materials which realize TCI phases,\cite{Kargarian2013, Wrasse2014, Zhou2018}
and studying the robustness of their boundary states.\cite{Morimoto2015}

One of the main interesting challenges is to construct new types of topological phases by exploiting the spatial symmetries of the system. In this context, layered structures of suitable materials can be engineered to build topologically nontrivial heterostructures.\cite{Trifunovic2016, Volpez2017} It has been shown that 3D TCIs can be constructed by stacking 2D TCI layers,\cite{Kim2015} but also by using 2D Chern insulators stacked in an antiferromagnetic fashion, such that the sign of the Chern number changes in every second layer.\cite{Mong2010} The latter model, called an ``antiferromagnetic topological insulator'', was recently modified in order to describe 3D TCIs protected by mirror symmetry,\cite{Fulga2016} glide symmetry,\cite{Kim2016, Lu2017}
to show that TCIs can occur in periodically driven systems,\cite{Ladovrechis2018}
as well as to study the newly discovered higher-order TIs.\cite{Schindler2018, Trifunovic2018, Khalaf2018, Kooi2018}

From an experimental point of view, building a heterostructure of Chern insulators with opposite topological invariants is hindered by an immediate practical difficulty. To change the sign of the Chern number one must typically reverse the direction of the applied magnetic field. While this may be achieved on sufficiently long distances, a field reversal on the atomic scale of the heterostructure's unit cell is highly impractical. One way to overcome this difficulty would be to use 3D materials which order anti-ferromagnetically and simultaneously realize quantum anomalous Hall phases in the 2D limit. However, to our knowledge such a material has not yet been reported.

\begin{figure}[tb]
 \includegraphics[width=\columnwidth]{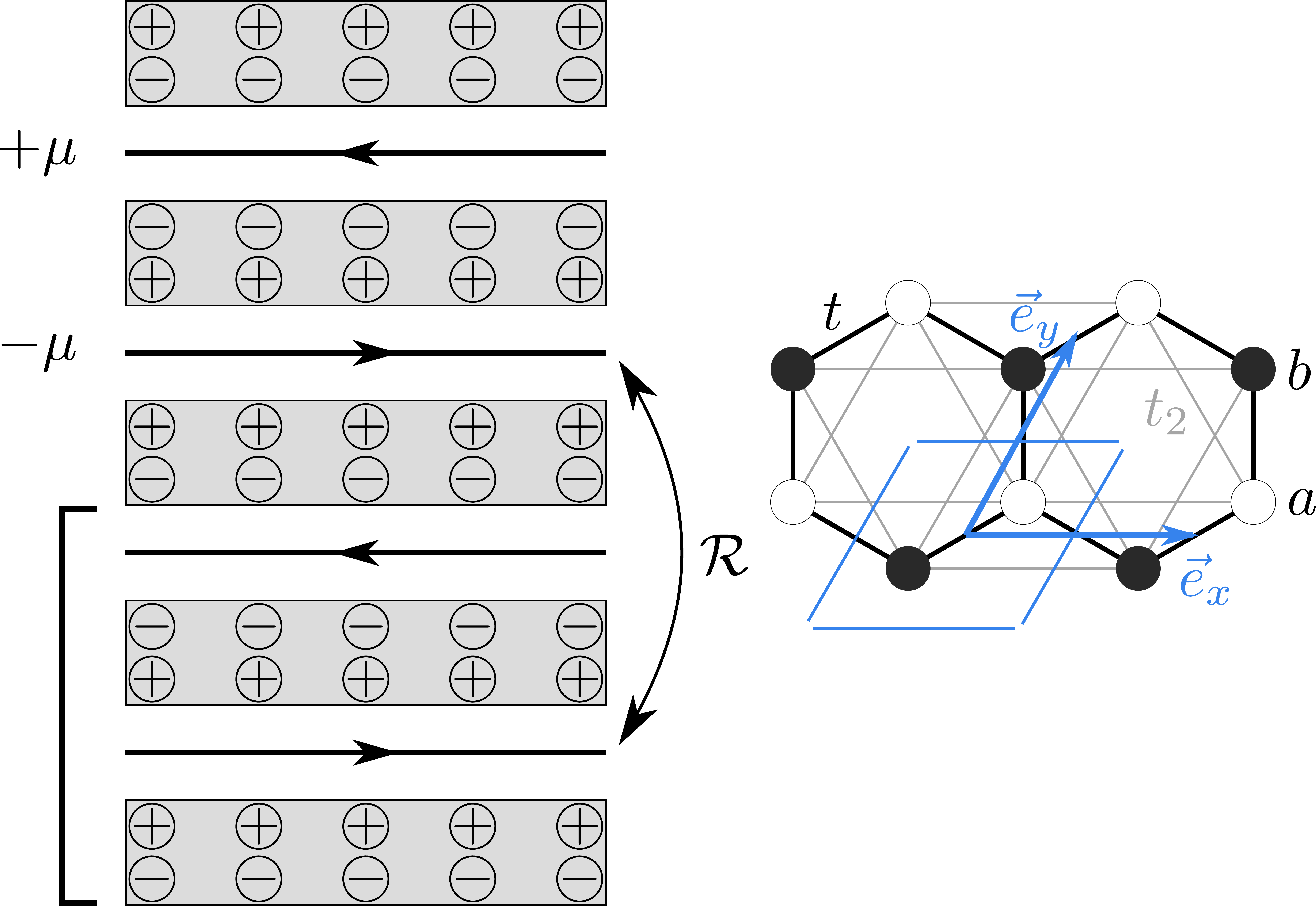}
 \caption{Left: Three-dimensional system formed out of graphene layers (horizontal lines) separated by thin insulating layers (gray boxes). The spacers are polar, having a positively charged ($+$) and a negatively charged ($-$) side. Using spacer layers with an alternating orientation leads to graphene sheets which have an alternating doping ($\pm\mu$). The unit cell of the heterostructure (bracket) consists of two graphene layers, and the full system shows reflection symmetry about one layer (${\cal R}$). By applying a uniform magnetic field along the stacking direction, neighboring graphene layers form quantum Hall phases with opposite Chern numbers, such that their chiral edge states propagate in opposite directions (horizontal arrows). Right: our conventions for the graphene lattice, with Bravais vectors $\vec{e}_x$ and $\vec{e}_y$. Nearest and next nearest neighbor hoppings are labeled $t$ and $t_2$. There are two sites, denoted $a$ and $b$ in every unit cell (marked by a blue contour).\label{fig:stack}}
\end{figure}

In this work, we adopt an entirely different strategy, one which does not rely on alternating magnetic fields, but on the Dirac nature of charge carriers in graphene. It is well known that, due to the zeroth Landau level of graphene, the velocity of the quantum Hall edge states can be switched not only by reversing the magnetic field direction, but also under a constant field, by a small shift of the chemical potential across the charge neutrality point.\cite{Zhang2005}
As such, we consider a heterostructure in which the graphene layers are separated by 2D insulating systems which are polar, as shown in Fig.~\ref{fig:stack}. By reversing the polarization of every second spacer layer, it is in principle possible to obtain a system in which adjacent graphene sheets have an alternating electron and hole doping. In this case, applying a uniform magnetic field along the stacking direction opens a topological gap in the graphene layers, but in such a way that they carry opposite Chern numbers.

In the following, we examine the system in two different limits, depending on which materials are used for the spacer layers. If the latter are composed of light elements, we can expect spin-orbit coupling (SOC) to be negligible, and the heterostructure can be treated as an effectively spinless model. In Section \ref{sec:QHE}, we show that in this case an intrinsically magnetic TCI phase is realized, one which requires an externally applied magnetic field to exist. On the other hand, when the polar spacers contain heavy elements, they may lead to proximity-induced SOC in the graphene layers, such that each layer forms a quantum-spin Hall phase.\cite{Kane2005a} In Section \ref{sec:QSHE}, we show that when SOC terms are larger than the doping, a time-reversal symmetric TCI phase can be realized. We conclude and discuss directions for future research in Section \ref{sec:conc}.

\section{Stack of Chern insulating layers of graphene}
\label{sec:QHE}

We begin by examining the first of two limits, in which the graphene sheets experience a negligible SOC, such that the heterostructure forms an effectively spinless system.
In this limit, we show that due to the alternating electron and hole doping of adjacent layers, applying a magnetic field parallel to the stacking direction results in a mirror symmetry protected TCI.

In the absence of SOC, the out of plane spin component of electrons in graphene is conserved, such that each spin sector can be treated independently. We therefore model the heterostructure as a 3D system of spinless electrons hopping on a lattice of AA-stacked honeycomb layers. The real space Hamiltonian reads
\begin{equation}\label{eq:H_nospin}
\begin{split}
\mathcal{H} = & \sum_{\left< ij \right>,\alpha}t\,{c^\dag_{i,\alpha}}{c^\pd_{j,\alpha}}
+\mu\sum_{i,\alpha}(-1)^\alpha{c^\dag_{i,\alpha}}{c^\pd_{i,\alpha}} \\
 & +\sum_{i,\alpha}\left[ t_z\,{c^\dag_{i,\alpha}}{c^\pd_{i,\alpha+1}}+{\rm h.c} \right],
\end{split}
\end{equation}
where $c^\dag_{i,\alpha}$ ($c^\pd_{i,\alpha}$) creates (annihilates) fermions on site $i$ in layer $\alpha$ and $\left<\ldots \right>$ denotes nearest neighbors (see Fig.~\ref{fig:stack}). The first term is a nearest neighbor hopping, which we set to $t=1$ throughout the following, whereas $\mu$ is an on-site energy which models the alternating doping of adjacent graphene systems. As such, there are two layers in each unit cell.
The last term of Eq.~\eqref{eq:H_nospin} models inter-layer coupling, with hopping to the layer below having an amplitude $t_z$ and hopping to the layer above an amplitude $t_z^*$. In practice, this term will decay exponentially with the separation of neighboring graphene sheets, requiring the use of very thin spacers. However, as we show in the following, a TCI phase can be realized even when $t_z$ is the smallest energy scale of the problem, provided it does not vanish exactly. In the latter case, the system cannot be treated as three-dimensional, since the heterostructure is composed of isolated 2D systems.

The momentum space form of Eq.~\eqref{eq:H_nospin} is given by 
\begin{equation}\label{eq:Hk_nospin}
\begin{split}
 {\cal H}(\vec{k}) &= \begin{pmatrix}
                      {\cal H}_+(k_x, k_y) & t_z^* + t_z e^{-i k_z} \\
                      t_z + t_z^* e^{i k_z} & {\cal H}_-(k_x, k_y)
                     \end{pmatrix} \\
  {\cal H}_\pm(k_x, k_y) &= t \big[ 1+\cos(k_x)+\cos(k_y) \big] \tau_x  \\
  & + t \big[ \sin(k_x) + \sin(k_y) \big] \tau_y \pm \mu\\
\end{split}
\end{equation}

Here, ${\cal H}_\pm$ are the Hamiltonians of graphene layers experiencing a $\pm\mu$ energy shift, $\vec{k}=(k_x,k_y,k_z)$, $k_{x,y}$ are the in-plane momentum components along $\vec{e}_{x,y}$ (see Fig.~\ref{fig:stack}), and $k_z$ is the momentum along the stacking direction. The Pauli matrices $\tau$ parameterize the $a$ and $b$ sublattice degree of freedom. Lastly, the $2\times 2$ grading on the first line of Eq.~\eqref{eq:Hk_nospin} encodes the degree of freedom associated to the two layers in the unit cell, which we denote in the following using Pauli matrices $\eta$.

Choosing a real valued inter-layer coupling, $t_z=t_z^*$, the Hamiltonian Eq.~\eqref{eq:Hk_nospin} obeys a spinless mirror symmetry of the form
\begin{equation}\label{eq:r_nospin}
 {\cal R}(k_z) = \tau_0 \otimes \begin{pmatrix}
                                 1 & 0 \\
                                 0 & e^{i k_z}
                                \end{pmatrix},
\end{equation}
such that
\begin{equation}\label{eq:Hrsym}
\mathcal{R}(k_z)\mathcal{H}(k_x,k_y,k_z)\mathcal{R}(k_z)^{-1} = \mathcal{H}(k_x,k_y,-k_z).
\end{equation}
As a consequence, the terms proportional to $t_z$ vanish on the mirror invariant plane of the Brillouin zone, $k_z=\pi$, and the two graphene monolayers are effectively decoupled from each other. Further, since for $k_z=\pi$ the mirror operator is ${\cal R}=\tau_0\eta_z$, electronic states in adjacent monolayers have different mirror eigenvalues, $+1$ and $-1$.
This naturally opens the possibility of stabilizing a mirror symmetry protected TCI if the graphene sheets enter Chern insulating phases when a magnetic field is applied. 

Since we are dealing with an effective spinless model valid for each of the two spin sectors, we introduce an orbital magnetic field through the usual Peierls substitution. We choose a gauge in which the in-plane hopping within each unit cell is modified as $t\to t\exp(i \Phi n_y)$, where $n_y$ is an integer labeling the unit cells in the $\vec{e}_y$ direction (see Fig.~\ref{fig:stack}), and $\Phi$ is the Peierls phase. The latter physically represents the number of magnetic fluxes penetrating a hexagonal plaquette with area $a_{\hexagon}$ due to a perpendicular magnetic field $B$, such that $\Phi = Ba_{\hexagon}e/h$.

\begin{figure}[tb]
 \includegraphics[width=\columnwidth]{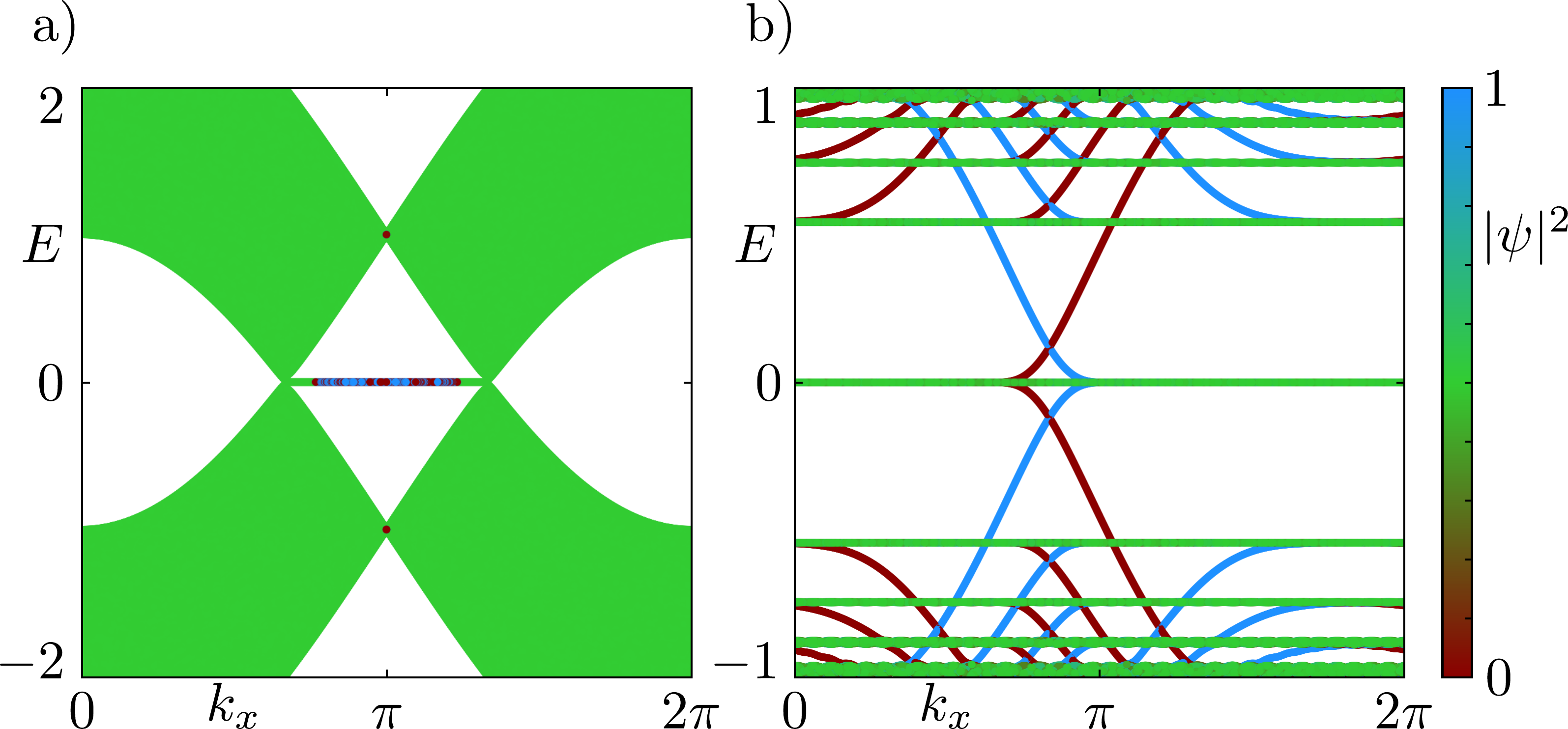}
 \caption{Bandstructure of a single monolayer of spinless graphene in a ribbon geometry (infinite along $\vec{e}_x$, $W=100$ unit cells along $\vec{e}_y$), using $t=1$ and $\mu=0$. In the absence of a magnetic field ($\Phi=0$, panel a), two bulk Dirac cones are connected by dispersionless boundary states localized on the two zig-zag edges of the ribbon. With a magnetic flux $\Phi=0.18$ (panel b) 
the bulk spectrum consists of Landau levels, and chiral edge modes appear at the two boundaries of the ribbon. The color scale denotes the probability density of a state integrated over half of the ribbon (unit cells indexed by $0\leq n_y < W/2$), such that modes localized on opposite boundaries of the ribbon are shown in blue and red, respectively.\label{fig:monolayer}}
\end{figure}

Using the Kwant code,\cite{Groth2014, SuppMat}
we compute the bandstructure of a single graphene sheet in a ribbon geometry with zig-zag edges, translationally invariant along $\vec{e}_x$, and consisting of $W=100$ unit cells in the $\vec{e}_y$ direction (see Fig.~\ref{fig:stack}). Note that our gauge choice for the Peierls substitution is only compatible with translation symmetry along $\vec{e}_x$.
As shown in Fig.~\ref{fig:monolayer}, for a single graphene monolayer the gapless Dirac cone spectrum becomes gapped under the addition of the orbital field, which leads to the formation of Landau levels. Characteristic to graphene and other hexagonal lattice systems, there exists a Landau level at the charge neutrality point, $E=0$. Away from this point, the system enters quantum Hall phases with opposite Chern numbers, $C=+1$ for $E>0$ and $C=-1$ for $E<0$.

Given the bandstructure of Fig.~\ref{fig:monolayer}b, we expect the alternating doping $\mu$ to ensure that adjacent graphene layers of the 3D system have opposite Chern numbers after the magnetic field is turned on, so that their chiral edge states have opposite velocities. Moreover, since states of neighboring graphene sheets are decoupled at $k_z=\pi$ and have opposite mirror eigenvalues, these chiral modes remain orthogonal on the mirror invariant plane due to Eq.~\eqref{eq:Hrsym}, leading to the formation of surface Dirac cones. The full heterostructure then realizes a mirror symmetric TCI phase with a mirror Chern number
\begin{equation}\label{eq:mirrorC}
 C_M = \frac{C_+-C_-}{2},
\end{equation}
where $C_\pm=\pm1$ are the Chern numbers (computed at $k_z=\pi$) of graphene layers experiencing a $\pm\mu$ energy shift, such that $C_M=1$. Note that, due to Eqs.~\eqref{eq:r_nospin} and \eqref{eq:Hrsym}, on the plane $k_z=0$ the reflection operator is equal to the identity matrix, ${\cal R}(0)=1$. The system then cannot be block-diagonalized into different mirror eigenspaces. It could still be possible that the full Chern number at $k_z=0$ is nonzero, but, lacking disjoint sectors with opposite mirror eigenvalue, such a topological phase would not be protected by mirror symmetry, corresponding instead to a stack of quantum Hall systems with co-propagating edge modes.

\begin{figure}[tb]
\begin{center}
 \includegraphics[width=\columnwidth]{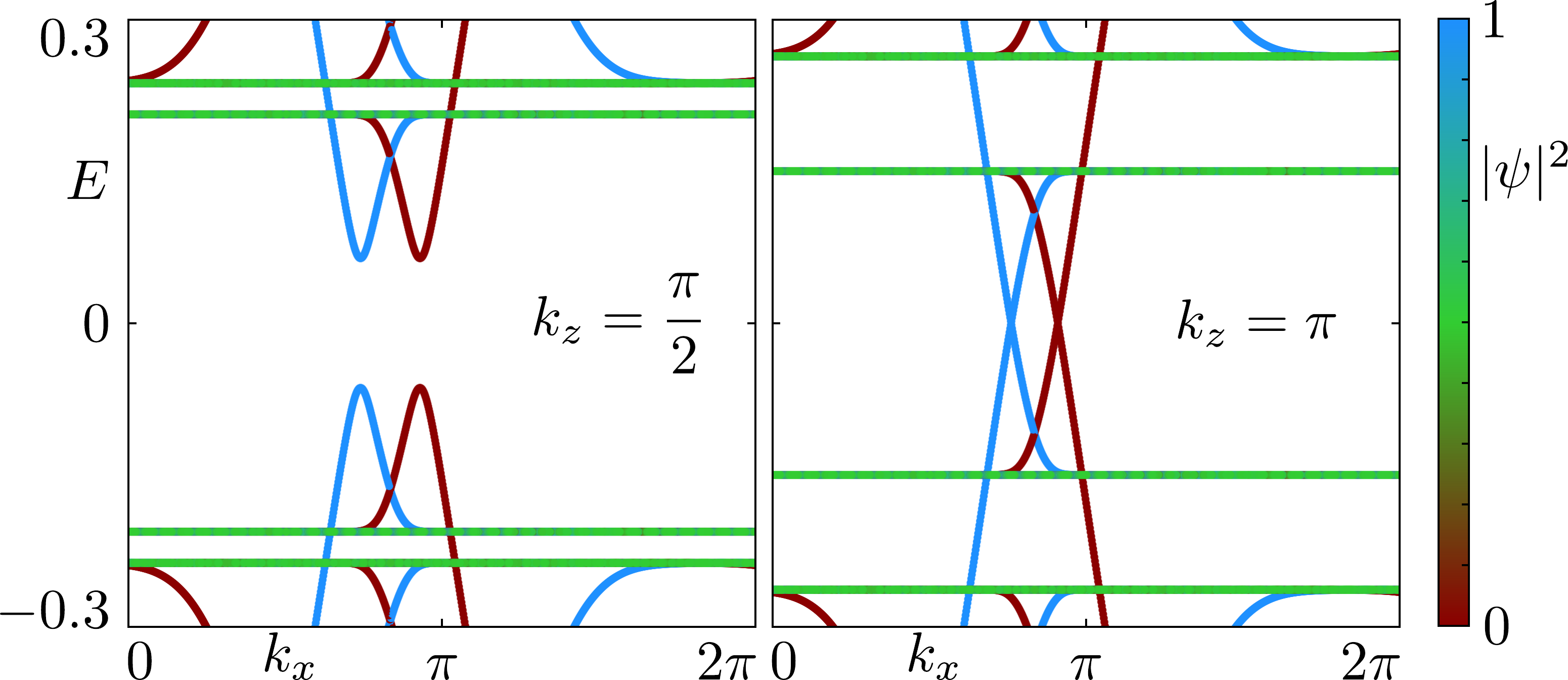}
 \caption{Bandstructure of the graphene heterostructure with Hamiltonian Eq.\eqref{eq:Hk_nospin} in an infinite slab geometry with hard wall boundary conditions in the $\vec{e}_y$ direction and a width of $W=100$ unit cells. We use $t=1$, $\Phi=0.18$, $\mu=0.3$, and $t_z=0.1$. The left and right panels show the bandstructures for $k_z=\pi/2$ and $k_z=\pi$, respectively. One Dirac cone appears on each surface, positioned on the mirror invariant $k_z=\pi$ line of the surface BZ. The color scale is the same as in Fig.~\ref{fig:monolayer}. In order for the inter-layer coupling to efficiently gap out the edge modes away from the mirror line, we have added a sublattice symmetry breaking term to the model $\mu_s\tau_z\eta_z$, with $\mu_s=0.15$. \label{fig:qhetci}}
\end{center}
\end{figure}

We confirm the presence of surface Dirac cones by computing the bandstructure of Eq.~\eqref{eq:Hk_nospin} in a slab geometry, infinite in the stacking direction and along $\vec{e}_x$, but containing $W=100$ sites in the $\vec{e}_y$ direction. As shown in Fig.~\ref{fig:qhetci}, on each surface the chiral modes of adjacent layers cross at $k_z=\pi$, but gap out away from this line, forming a surface Dirac cone protected by mirror symmetry. When determining the bandstructures, we have noticed that due to the high symmetry of ${\cal H}(\vec{k})$, the inter-layer coupling $t_z$ does not efficiently couple the chiral modes away from the mirror plane, leading to surface nodal lines that wind across the surface BZ in the $k_z$ direction. The surface nodal lines are a consequence of a spurious sublattice symmetry of the model, and occur both for zig-zag and armchair terminations of the graphene layers.
Since we are interested in the phenomenology of TCIs protected purely by mirror, we have
lowered the symmetry of the initial Hamiltonian ${\cal H}(\vec{k})$ by adding a sublattice symmetry breaking term, $\mu_s\tau_z\eta_z$, which enables the chiral edge modes to couple away from $k_z=\pi$. This term does not break the mirror symmetry Eq.~\eqref{eq:r_nospin}, such that the mirror Chern number remains non-trivial, provided that $\mu_s$ is not large enough to close the bulk gap.

As we have shown, the heterostructure of oppositely doped graphene layers enters a TCI phase under an externally applied magnetic field. Unlike previously observed TCIs, this phase is only present when time-reversal symmetry is explicitly broken, since the spectrum is gapless in the zero field case. As long as SOC is negligibly weak, each spin component of the graphene charge carriers behaves according to the Hamiltonian Eq.~\eqref{eq:Hk_nospin}, such that the full system contains two surface Dirac cones, which are protected by the conservation of the out of plane spin component.
Further, the precise form of the mirror symmetry Eq.~\eqref{eq:r_nospin} may be tuned by altering the materials forming the polar spacer layers, and therefore the inter-layer coupling $t_z$. If, for instance, we choose an imaginary hopping between graphene monolayers, $t_z=-t_z^*$, then the mirror operator would read ${\cal R}=\tau_0\otimes{\rm diag}(1,-e^{ik_z})$, and the surface Dirac cones would be positioned at a different mirror invariant plane, $k_z=0$. Notice however that for a generic, complex valued $t_z$ the heterostructure Hamiltonian Eq.~\eqref{eq:Hk_nospin} would break both this mirror symmetry and that of Eq.~\eqref{eq:r_nospin}. To introduce complex inter-layer hoppings one would have to modify Eq.~\eqref{eq:Hk_nospin} such that the phase of the hopping to the layer above is opposite to the phase of the hopping to the layer below. For instance, replacing the off-diagonal blocks of this Hamiltonian with $|t_z|e^{i \theta} (1+e^{i k_z})$ would preserve mirror symmetry for any value of the complex phase $\theta$, as evidenced by the fact that the term still vanishes at $k_z=\pi$.

Finally, we note that there is no threshold value of $t_z$ for which a TCI phase is realized, meaning that the inter-layer coupling can be the smallest energy scale of the problem. Reducing the value of $t_z$ by increasing the thickness of the spacer layers does not remove the topologically protected surface Dirac cones, but simply reduces their velocity in the $k_z$ direction.

\section{Stack of quantum spin-Hall layers of graphene}
\label{sec:QSHE}

The negligibly small value of SOC in free standing graphene\cite{Min2006, Yao2007} enabled us to use a spinless model when discussing the heterostructure of Fig.~\ref{fig:stack}, provided that the polar spacers contain light elements. It is however known that graphene in proximity to heavy atoms or 2D materials containing heavy atoms may lead to large induced SOC terms.\cite{Avsar2014, Kou2015}
In Ref.~\onlinecite{Kou2014} for instance, it was shown that a SOC-driven quantum spin-Hall phase with a gap as large as $80$ meV may be realized in graphene sandwiched between oppositely oriented 2D layers of BiTeX (X=Cl, Br, I). Motivated by this fact, in the following we study the heterostructure in the limit in which SOC is larger than the alternating doping of adjacent graphene sheets.

We describe the system using AA-stacked copies of spin--$\frac{1}{2}$ graphene models. The 3D real space Hamiltonian now reads
\begin{equation} \label{eq:graphene_stack}
\begin{split}
\mathcal{H}_{\frac{1}{2}} & = \sum_{\left< ij \right>,\alpha}t\,{c^\dag_{i,\alpha}}{c^\pd_{j,\alpha}}
+\mu\sum_{i,\alpha}(-1)^\alpha{c^\dag_{i,\alpha}}{c^\pd_{i,\alpha}}
   \\
 & + \sum_{\left< \left< ij \right> \right>,\alpha}it_2{\nu_{ij}}{c^\dag_{i,\alpha}}{\sigma_z}{c^\pd_{j,\alpha}} \\
 & +\sum_{i,\alpha}\left[ {c^\dag_{i,\alpha}}T_z{c^\pd_{i,\alpha+1}}+{\rm h.c} \right],
\end{split}
\end{equation}
where $c^\dag_{i,\alpha}=\left( c^\dag_{i,\alpha,\uparrow} c^\dag_{i,\alpha,\downarrow} \right)$ creates fermions with spin $\uparrow,\downarrow$ on site $i$ in layer $\alpha$, $\left<\ldots \right>$ and $\left< \left< \ldots \right> \right>$ denote nearest and next nearest neighbors (see Fig.~\ref{fig:stack}), and the Pauli matrices $\sigma$ parameterize the spin degree of freedom. The first two terms, $t$ and $\mu$, have the same meaning as before, whereas the term proportional to $t_2$ is the usual intrinsic SOC term,\cite{Kane2005a} an imaginary next nearest neighbor hopping. The sign $\nu_{ij}=\pm1$ is positive whenever the path connecting sites $i$ and $j$ rotates counter-clockwise, and negative for a clockwise rotation. Finally, $T_z$ is a matrix describing electron hopping between neighboring graphene layers.

As before, we begin by discussing the decoupled limit $T_z=0$, when each of the graphene layers is an independent 2D system. Since in the simple model Eq.~\eqref{eq:graphene_stack} the SOC term commutes with $\sigma_z$, we can write the Hamiltonian separately for each spin component $s=\pm$ and each of the two layers in a unit cell $l=\pm$ as
\begin{equation}\label{eq:Hpm}
\begin{split}
 {\cal H}_{l=\pm,s=\pm} = & t \big[ 1+\cos(k_x)+\cos(k_y) \big]\tau_x + \\
 & t \big[ \sin(k_x) + \sin(k_y) \big]\tau_y + l\cdot \mu\tau_0 + \\
 & s\cdot 2 t_2 \big[ \sin(k_x)-\sin(k_y) \\
 & -\sin(k_x-k_y)  \big] \tau_0.
\end{split}
\end{equation}

The heterostructure obeys a spinful time-reversal symmetry with operator ${\cal T} = i \tau_0\eta_0 \sigma_y{\cal K}$ and ${\cal K}$ complex conjugation. Further, the system also obeys a spinful mirror symmetry about one layer, which takes the form
\begin{equation}\label{eq:rsym}
 {\cal R}_{\frac{1}{2}}(k_z) = \begin{pmatrix}
             \tau_0\sigma_z & 0 \\
             0 & \tau_0\sigma_z e^{i k_z}
            \end{pmatrix},
\end{equation}
where the $2\times2$ grading is in the layer degree of freedom, $\eta$.
Note that the reflection symmetry Eq.~\eqref{eq:rsym} anti-commutes with the time-reversal symmetry operator. In general, the commutation relation between the two operators is gauge dependent, since it is always possible to re-define ${\cal R}_{\frac{1}{2}} \to i {\cal R}_{\frac{1}{2}}$, such that the new operator commutes with time-reversal. We choose the basis conventionally used in topological classification studies,\cite{Chiu2013} in which the two symmetries anti-commute if the system is spinful.

The two spin eigenstates in each monolayer have opposite mirror eigenvalues.\cite{Liu2013, Dominguez2018} This means that under the addition of an intrinsic SOC term, $t_2>0$, each graphene sheet simultaneously realizes a quantum spin-Hall phase as well as a 2D TCI phase, since the different spin sectors have opposite Chern numbers $C=\pm1$.

For the inter-layer coupling we choose a term which respects both time-reversal as well as mirror symmetry, but mixes the two spin components, as one can expect when the polar spacers contain heavy elements. We set $T_z=i\sigma_x t_z$ in Eq.~\eqref{eq:graphene_stack}, where the real number $t_z$ is the strength of the coupling, such that the full momentum space Hamiltonian is
\begin{equation}\label{eq:Hkspin}
 {\cal H}_{\frac{1}{2}}(\vec{k}) =
 \begin{pmatrix}
  {\cal H}_{+,+} & 0 & 0 & A \\
  0 & {\cal H}_{+,-} & A & 0 \\
  0 & A^\dag & {\cal H}_{-,+} & 0 \\
  A^\dag & 0 & 0 & {\cal H}_{-,-} \\
 \end{pmatrix}
\end{equation}
with $A=it_z(1-e^{i k_z})\tau_0$.
Notice that according to Eq.~\eqref{eq:rsym}, even when the inter-layer coupling is added, there are now two different planes on which a mirror Chern number can be defined, $k_z=0$ and $k_z=\pi$, unlike the spinless model discussed in the previous section. Crucially however, the mirror eigenvalues of every second layer reverse when going from $k_z=0$ to $k_z=\pi$, which allows for a different mirror Chern number on each mirror invariant plane. When $k_z=0$, eigenstates of the spin-up sector (i.e. those of ${\cal H}_{+,+}$ and ${\cal H}_{-,+}$) have the same mirror eigenvalue as well as the same Chern number, both of which are opposite to those of ${\cal H}_{+,-}$ and ${\cal H}_{-,-}$. As such, the 3D coupled system realizes a TCI with mirror Chern number $C_M=2$, and two surface Dirac cones are expected to appear on the $k_z=0$ line of the surface BZ. On the other plane, $k_z=\pi$, the mirror eigenvalues switch both when changing the spin sector as well as the layer, leading to a trivial topological invariant. This is because the eigenspace with positive mirror eigenvalue is formed by ${\cal H}_{+,+}$ and ${\cal H}_{-,-}$, which in total have a vanishing Chern number.

\begin{figure}[tb]
\begin{center}
 \includegraphics[width=\columnwidth]{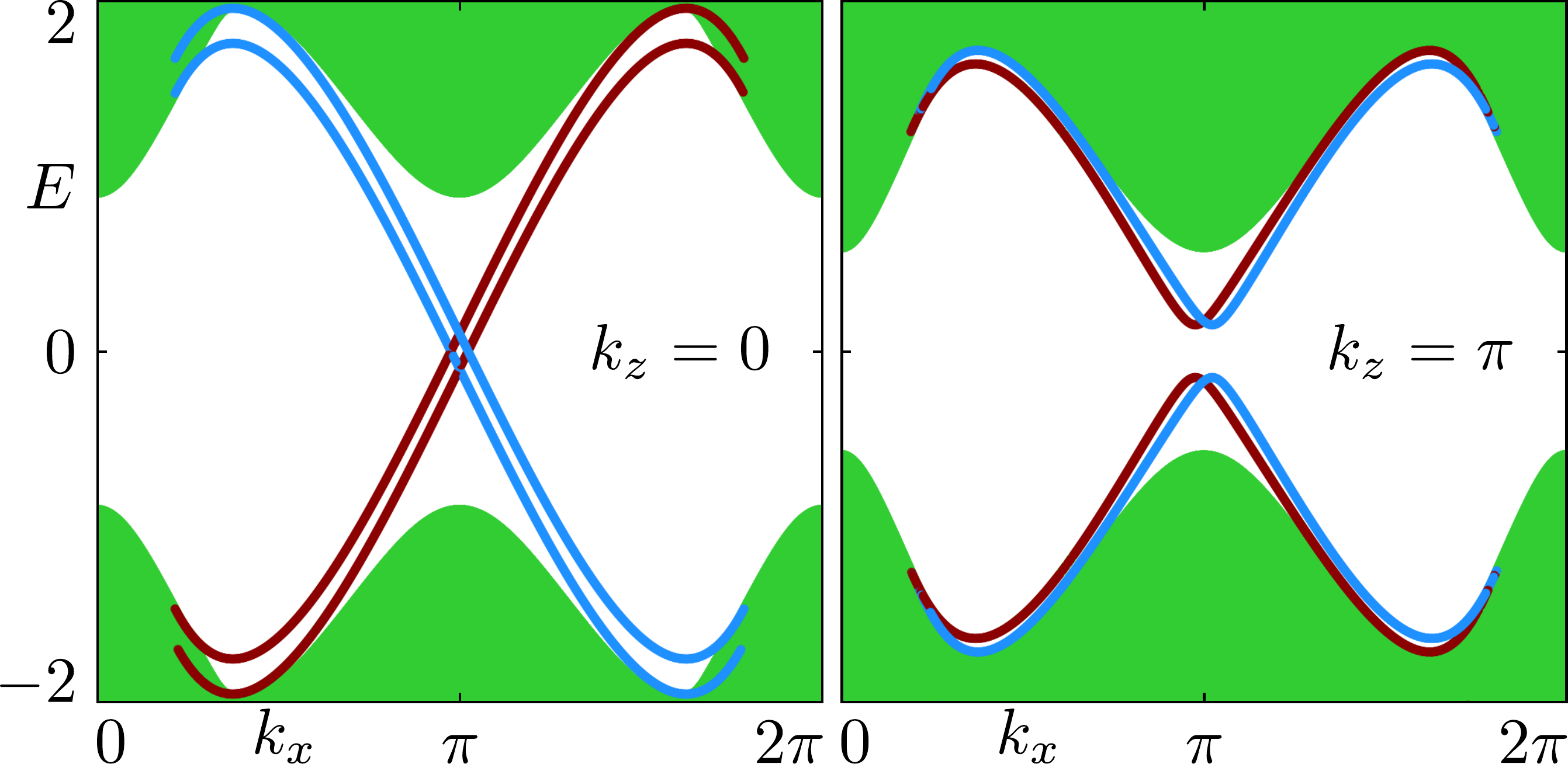}
 \caption{Bandstructure of the spinful graphene heterostructure [Eq.~\eqref{eq:Hkspin}] in an infinite slab geometry ($W=100$ unit cells along $\vec{e}_y$). We use $t=1$, $t_2=0.5$, $\mu=0.1$, and $t_z=0.2$. Only bulk modes (shown in green) and states on one of the two surfaces are plotted. The color of the surface modes denotes the mirror sector of each state: red for an eigenvalue $+1$ and blue for $-1$.
 At $k_z=0$ (left), the nonzero mirror Chern number leads to the appearance of two Dirac cones on the surface. States having the same mirror eigenvalue propagate in the same direction, so they are topologically protected. In contrast, for the other mirror invariant plane $k_z=\pi$ (right), the mirror Chern number vanishes. There are both left and right moving surface modes in each of the two mirror sectors, which are gapped out by the inter-layer coupling term. \label{fig:qshetci}}
\end{center}
\end{figure}

To confirm the presence of surface Dirac cones only at $k_z=0$, we plot in Fig.~\ref{fig:qshetci} the bandstructure of the system in an infinite slab geometry, with translational invariance along the stacking direction and $\vec{e}_x$, and containing $W=100$ unit cells in the $\vec{e}_y$ direction. The intrinsic SOC term $t_2=0.5$ is now larger than the alternating doping, $\mu=0.1$, such that each graphene layer independently realizes a quantum spin-Hall phase. At $k_z=0$, the mirror Chern number $C_M=2$ means that surface states with the same mirror eigenvalue propagate in the same direction, such that they cannot be gaped out. In contrast, at $k_z=\pi$, there are surface modes with opposite velocities in each mirror eigenspace, allowing the inter-layer coupling to produce a gapped surface.

Finally, notice that for this system the topological surface modes would persist even in the limit of vanishing doping, $\mu=0$. In this case, the unit cell would be halved, containing a single monolayer, and the heterostructure would realize a weak topological insulator, protected by time-reversal symmetry and translation along the stacking direction. However, the additional mirror symmetry Eq.~\eqref{eq:rsym} leads to an increased protection of the surface Dirac cones, allowing them to persist even as $\mu\neq0$, due to the system's non-trivial mirror Chern number.

\section{Conclusion}
\label{sec:conc}

We have shown that multilayers of graphene can exhibit a topological crystalline insulating phase protected by reflection symmetry. We considered a heterostructure formed by graphene monolayers sandwiched between oppositely oriented 2D polar spacers, such as BiTeX\cite{Kou2014} or ultra-thin ferroelectric polymers.\cite{Bune1998, Chu2006}
The spacers may lead both to an alternating doping as well as to a proximity induced SOC in the graphene sheets. Both limits were shown to lead to a mirror-symmetry protected TCI phase, hosting two Dirac cones on each surface. When the polar spacers are made of light elements, such that they induce a negligibly small SOC, the heterostructure can be treated as an effectively spinless system. In this case, we have shown that due to the alternating electron and hole doping of adjacent graphene layers, they form quantum Hall phases with opposite Chern numbers under a uniform magnetic field. The resulting phase is an ``intrinsically magnetic TCI'', one which requires the breaking of time-reversal symmetry in order to exist. In the opposite limit, when SOC is larger than the doping, the system instead realizes a time-reversal symmetric TCI with a mirror Chern number of 2. Similar to KHgX (X=As, Sb, Bi),\cite{Wang2016} the surface modes can be understood as originating from two quantum spin-Hall systems which are forbidden to gap each other out in the presence of mirror symmetry.

Our work focused only on toy models and discussed the possibility for TCI heterostructures to exist as a proof of principle. We hope that this study will motivate future ab initio approaches to graphene heterostructures and their potential for realizing TCIs. There are a large number of 2D materials which may be combined in van der Waals heterostructures\cite{Geim2013, Novoselov2016, Liu2016} and which show a variety of physical properties, such as polarity, magnetism, or SOC. It would be interesting to combine machine learning algorithms with density functional theory methods to automate the search for topologically non-trivial heterostructures.

On the experimental side, we expect that such layered systems will first be fabricated using only a few graphene sheets, so that the system is not fully three-dimensional. In the small thickness regime, it may be possible to gate the sample using external electrodes, such that the doping of adjacent graphene monolayers can be more readily controlled. Further, studying heterostructures composed of a few layers would open the possibility of observing the so called ``even-odd effect'' in TCIs. The latter was originally discussed in WTIs,\cite{Ringel2012} and states that a system containing an even number of layers may be gapped by inter-layer coupling, whereas one containing an odd number must host topologically protected gapless modes on its surface. For the systems studied here, the same criterion applies with respect to the parity of the number of graphene sheets, both in the time-reversal symmetric and in the magnetic TCI limits.

\emph{Note added:} In the final stages of writing this manuscript, we became aware of the similar proposal of Ref.~\onlinecite{Kooi2018}, which considers alternating electron and hole doping in stacked silicene layers in order to produce higher order TIs.

\acknowledgements

We thank Ady Stern for fruitful discussions.

\bibliography{bibfile}

\end{document}